# Photo-Chemically Directed Self-Assembly of Carbon Nanotubes on Surfaces

Zhaoying Hu[1], James B. Hannon[2], Hongsik Park[2], Shu-Jen Han[2], George S. Tulevski[2], Ali Afzali[2,a], Michael Liehr[1]

[1]*College of Nanoscale Science and Engineering, The State University of New York at Albany, Albany, New York 12203, USA*

[2]*IBM T. J. Watson Research Center, Yorktown Heights, New York 10598, United States*

Transistors incorporating single-wall carbon nanotubes (CNTs) as the channel material are used in a variety of electronics applications. However, a competitive CNT-based technology requires the precise placement of CNTs at predefined locations of a substrate. One promising placement approach is to use chemical recognition to bind CNTs from solution at the desired locations on a surface. Producing the chemical pattern on the substrate is challenging. Here we describe a one-step patterning approach based on a highly photosensitive surface monolayer. The monolayer contains chromophopric group as light sensitive body with heteroatoms as high quantum yield photolysis center. As deposited, the layer will bind CNTs from solution. However, when exposed to ultraviolet (UV) light with a low dose (60 mJ/cm$^2$) similar to that used for conventional photoresists, the monolayer cleaves and no longer binds CNTs. These features allow standard, wafer-scale UV lithography processes to be used to form a patterned chemical monolayer without the need for complex substrate patterning or monolayer stamping.

[a] **Author to whom correspondence should be addressed.  Electronic mail:** afzali@us.ibm.com.

Single-wall carbon nanotube (CNT) devices hold promise for potential applications[1] in high-performance logic,[2] flexible electronics,[3] physical unclonable function,[4] and biological or chemical sensing.[5] One difficulty in manufacturing CNT-based technology is that new fabrication techniques are required: CNTs must be incorporated into the target devices, and the nanoscale manipulation processes needed to accomplish this are in their infancy. Various approaches for manipulating CNTs have been demonstrated, including growth of aligned CNTs on quartz followed by transfer to the target substrate,[6] self-assembly by dielectrophoresis,[7] shear forces,[8] and chemical recognition.[9]

CNT placement from solution is an attractive approach for manufacturing CNT electronics. Most importantly, this approach separates the process of CNT purification (the elimination of metallic CNTs) from CNT placement. In addition, there are many solution-based mechanisms for placement that can be exploited. For example, in several studies[10,11] patterns with hydrophilic and hydrophobic regions were used to guide CNT placement. Although these methods show very good selectivity and high CNT density, "bundling" of the CNTs was observed because the solutions were allowed to dry on the surface. In later work, oxide surfaces with $SiO_2$ and $HfO_2$ regions were used to define surface patterns. Specially modified CNTs, containing acidic functional groups, will adhere to the $HfO_2$ regions, but do not bond to the $SiO_2$ regions.[12] Recently, acidic functional groups is formed as a monolayer on the $HfO_2$ regions instead of functionalizing CNTs.[9,13] In this approach, the monolayer contains a positively charged functional group that interacts with the negatively-charged surfactant wrapping the CNTs. However, the use of patterned oxide substrates to imprint a pattern on the self-assembled monolayer make the fabrication process complicated.

Alternative methods of patterning self-assembly monolayers have been extensively investigated. For example, scanning probe lithography,[14] soft lithography,[15] electron beam lithography,[16] scanning near-field photolithography[17] and UV photolithography[18] have all been used. Most of these methods are 'serial,' and are not compatible with large-scale, high-throughput manufacturing. UV photolithography is the most common patterning approach used in semiconductor manufacturing, which makes it a natural choice for patterning self-assembled monolayers. Indeed, patterning of photosensitive monolayers has been demonstrated for protein immobilization[19,20] and also CNT placement.[18] However, the required doses are in the range of 1J/cm$^2$ [18,21] which is almost two orders of magnitude larger than the doses used for commercial photoresists, making these approaches inefficient.

The desired properties of a charged photosensitive monolayer for CNT placement are: (1) the formation of a self-assembly monolayer on the $HfO_2$ for CNT placement directly on the gate dielectric; (2) cleavage of the molecule upon UV light exposure; (3) a significant change in the interaction between the molecule and a CNT upon cleavage, giving rise to selective placement.

Here we describe a charged photosensitive monolayer compatible with conventional photolithography for carbon nanotube placement, but which requires a far lower dose for patterning: about 60 mJ/cm$^2$. The photosensitive compound used in this study is 4-(hydroxycarbamoyl)-1-((2-nitrobenzyl)oxy)pyridin-1-ium (2-NBO), which is synthesized from commercial available compounds. Good CNT density and adsorption selectivity have been demonstrated. CNTs from solution bind to the as-deposited monolayer, but do not bind to regions exposed to UV radiation. It is the most UV-sensitive monolayer synthesized to date.

In Fig. 1(a), the UV-vis absorption measurement (UV-vis-NIR absorption spectrometer, Perkin-Elmer Lamda 950) of the photosensitive compound solution shows a rapid increase in absorption for wavelength smaller than 300 nm and a small peak at 255 nm. The absorption peak is close to 248 nm which is used in standard UV lithography tools. Photosensitive monolayers were prepared immersing a plasma cleaned 10-nm $HfO_2$ film on Si substrate into a 10 mM solution of 2-NBO dissolved in a mixed solvent (methanol:ethanol:water=7:2:1) for 1 h, then rinsing with ethanol and drying with $N_2$. Upon exposure to 248 nm UV light, the monolayer compound is excited and fragments through the N-O bond, yielding a neutral surface, as shown in Fig. 1(b).

The kinetics of the bond cleavage can be determined by measuring the water contact angle (VCA optima instrument), and IR absorption spectrum (Thermo Nicolet Nexus IR spectrometer), as function of UV dose. The UV light source is a KrF laser with 248 nm wavelength. As shown in Fig.1(c), initially the water contact angle is 42°. Upon exposure, the contact angle decreases monotonically, and saturates at 16° after UV exposure with dose of 60 $mJ/cm^2$. Thus we infer the optimal exposure dose would be between 30 to 60 $mJ/cm^2$. IR absorption spectroscopy suggest that the change in the contact angle is associated with the removal of the $NO_2$ group from the 2-NBO. As shown in Fig. 1(d), a thick 2-NBO film exhibits two strong absorption peaks at 1346 $cm^{-1}$ and 1531 $cm^{-1}$ which arise from the two asymmetric stretching modes of $NO_2$ group. These two peaks are present for the 2-NBO monolayer, but their intensity decreases with UV exposure. We conclude that UV exposure leads to the removal of the $NO_2$ group. The $NO_2$ group is not essential for the photosensitivity but the N-O heteroatoms is, as observed in the six compounds screened in our experiments (not shown).

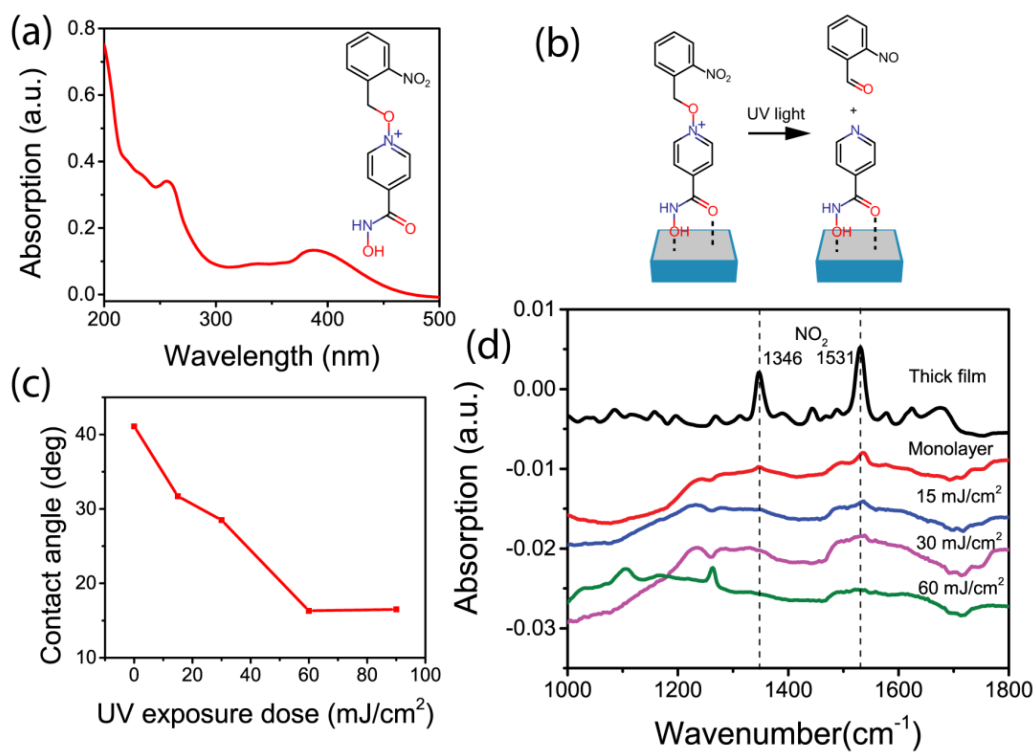

FIG. 1. Kinetics of the charged photosensitive monolayer upon UV light exposure. (a) UV-vis absorption spectroscopy of 2-NBO solution. (b) Photoinitiated bond cleavage mechanism. (c) Water contact angle of the monolayer after different UV exposure dose. (d) IR absorption spectral of the thick film and the monolayer after different exposure dose.

Patterning of the 2-NBO monolayer was carried out in a home-made UV exposure system by placing a glass shadow mask directly on top of the substrate. Following UV exposure, the monolayer was rinsed by ethanol for 5 s. A high contrast pattern was observed between the un-exposed region (dark) and the exposed region (bright) in scanning electron microscopy (SEM) images (Fig. 2(b)). The contrast is due to differences in secondary electron emission or surface work function between areas with pristine and those with cleaved molecules. As shown in Fig. 2(c), the pattern transferred from the mask to the monolayer is accurate according to the intensity profile obtained from Fig. 2(b).

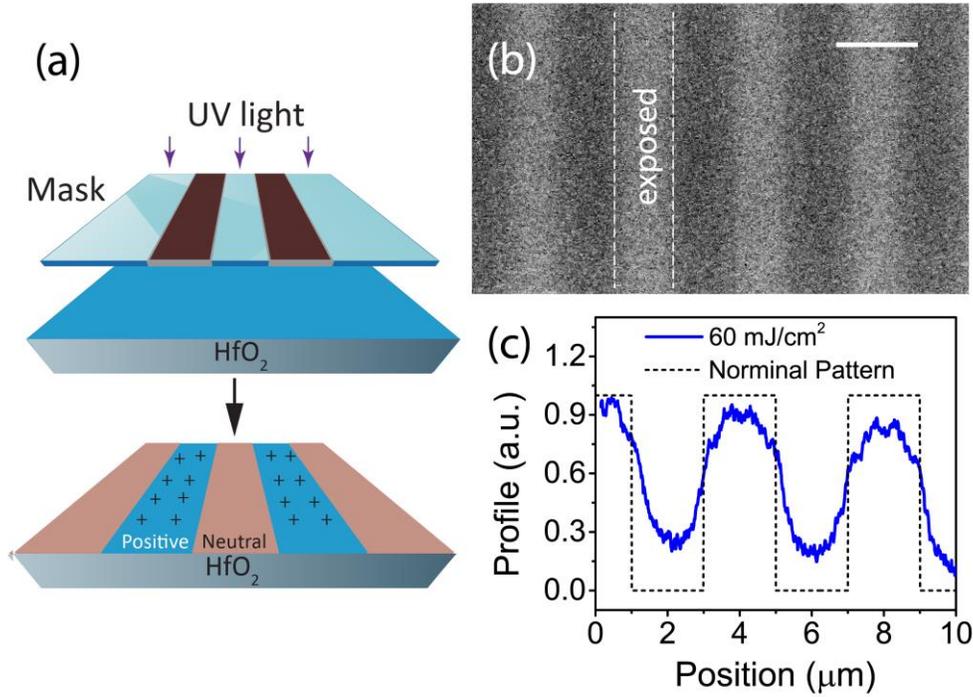

FIG. 2. Monolayer patterning by contact photolithography. (a) Schematic of contact photolithography using 248 nm wavelength laser. (b) SEM images (scale bar: 2 μm) of patterned monolayer on HfO$_2$ using dose of 60 mJ/cm$^2$. (c) Intensity Profile of (b).

In order to better understand the surface properties, frequency-modulated (FM)-Kelvin probe force microscopy (KPFM) study was carried out on the UV patterned monolayer sample out in ambient environment using Bruker Dimension Icon SPM. An atomic force microscopy (AFM) image is obtained by the PeakForce tapping mode. A contact potential difference (CPD) $V_{CPD}$ map is generated by applying a DC voltage V$_b$ to nullify the electrostatic force $F_{es}$ between the tip and the substrate. When the external DC bias V$_b$ is applied through the tip with the sample grounded, V$_{CPD}$ is equal to -V$_b$. As shown in the band diagram for the Kelvin probe force measurement (Fig. 3(a)), V$_{CPD}$ is given by $V_{CPD} = -V_b = -(\Phi_{Tip} - \Phi_{Sample} - Q_S/C_{ox})$. Assuming the work function of the tip and the HfO$_2$/Si stack do not change over the scanning area, the difference of $V_{CPD}$ is thus only determined by the surface charge $Q_S$. Larger values of $V_{CPD}$ indicate a more positive charge on the surface. The topography image (Fig. 3(b)) shows poor contrast due to the small height difference of the monolayers. However, Fig. 3(c) shows good contrasts for the potential map of the UV patterned monolayer, on which the purple region is unexposed and the green region is exposed. The unexposed region shows a higher $V_{CPD}$ than the value of the exposed region by ~75 mV, as shown in Fig. 3(d) and Fig. 3(e). The symmetric potential histogram in Fig. 3(e) shows the areas of these two regions are the same, which agrees with the SEM image intensity profile. The measured surface potential was reported to correlate with the isoelectric point (IEP).[22] Thus we can infer that the unexposed region with higher surface potential has larger isoelectric point than the exposed region. This conclusion also agrees with the theoretical calculated values for the 2-NBO molecule before (IEP=13.28) and after UV cleavage (IEP=6.29). Namely, the unexposed region is positively charged in a neutral aqueous solution but the exposed region mostly remains neutral.

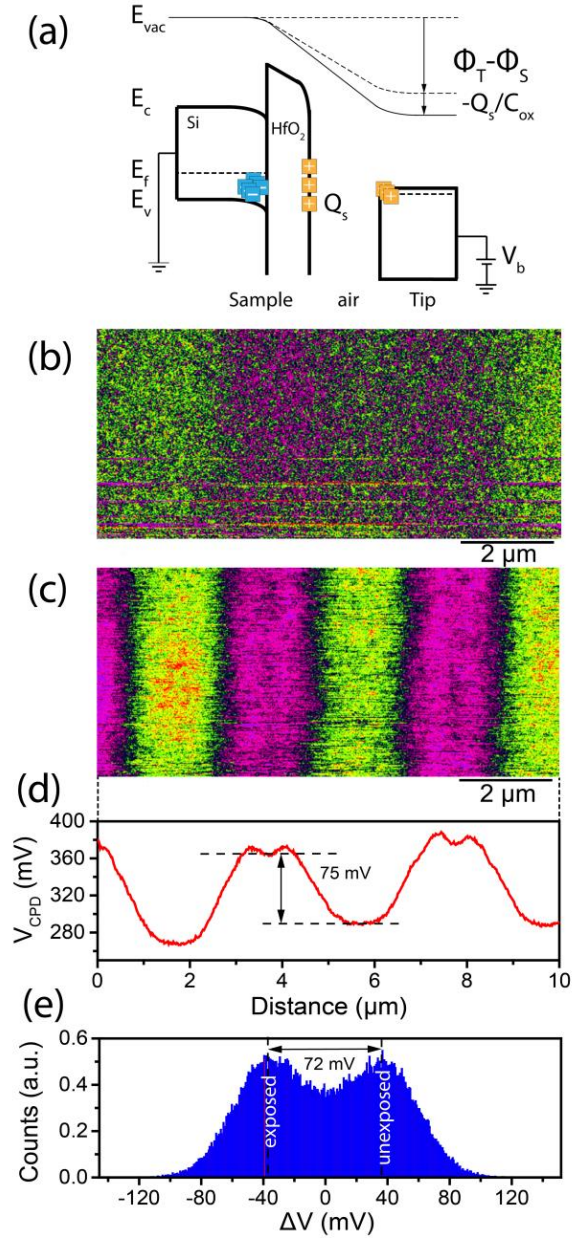

FIG. 3. A KPFM study of the UV light patterned photosensitive monolayer on $HfO_2$. (a) Schematic band diagram of Kelvin probe force measurements used to characterize surface potential. (b) AFM image and (c) Surface potential (KPFM) image of patterned monolayer on $HfO_2$ substrate with UV dose of 60 mJ/cm$^2$. (d) Surface potential profile showing a potential difference of ~75 mV between the exposed and the unexposed area. (e) Potential distribution histogram averaged over the image (c).

Carbon nanotube placement was achieved by ion-exchange chemistry, in a manner similar to that described in Ref 8.[9] A schematic illustration of selective carbon nanotube placement is shown in Fig. 4(a). In an aqueous solution, the 2-NBO layer will be positively charged ($N^+$) with a double layer formed from the $PF_6^-$ counter ions. The anionic surfactant wrapped CNT is negatively charged, surrounded by positive counter ions ($Na^+$). The strong Coulombic attraction force generated by the gradients of the counter ions leads to binding of CNT to monolayer and releasing of $NaPF_6$ salt into solution. In contrast, ion-exchange reaction is prohibited on the cleaved (neutral) monolayer, and surfactant-wrapped CNTs will not bind. We note that, in principle, this monolayer can be used to selectively bind any negatively-charged nanoparticle from solution.

An aqueous CNT solution was prepared using the anionic surfactant sodium dodecyl sulphate, (SDS). Excess surfactant in the CNT solution was removed by 4 cycles of dialysis in deionized (DI) water. Following exposure to CNTs, via drop casting method, selective placement of CNTs was shown in Figs. 4(b)-(d). High density deposition of CNTs with a density of 60 – 75 CNTs/$\mu m^2$ on the unexposed region (dark) was achieved with no degradation in selectivity. Both good density and selectivity can be obtained for 2-μm lines, which is the minimum feature on our mask. The good selectivity is attributed to the high charge contrast between the exposed and the unexposed region as shown in Fig. 3(c). As shown in Fig. 4(d), our deposition method ensures most of the CNTs remain single tubes rather than lots of bundles seen in the methods using surface hydrophobicity.[18]

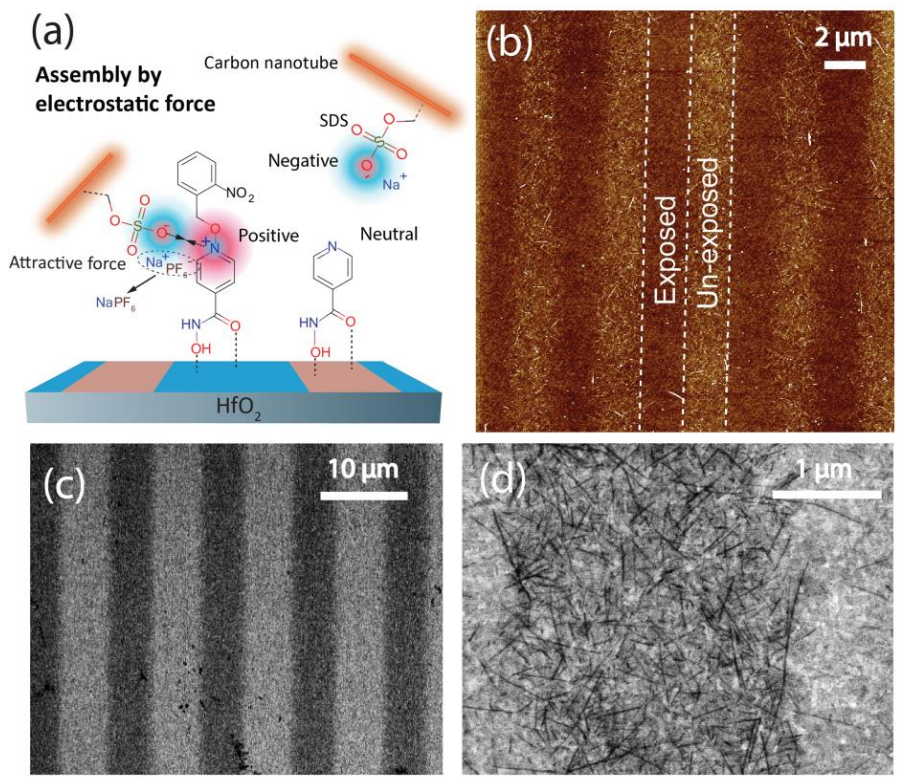

FIG. 4. Selective carbon nanotube placement by charged photosensitive monolayer. (a) Schematic illustration of carbon nanotube placement by photo-patterned charged monolayer. (b) AFM image, (c) and (d) SEM images of the nanotubes selectively deposited on UV light patterned monolayer on HfO$_2$ substrate by contact photolithography.

In summary, we have synthesized a photosensitive surface monolayer that can be used to pattern CNT deposition onto metal oxide surface from solution. We have demonstrated good selectivity and CNT density in 2-μm patterns generated using a photo mask. Water contact angle measurements and IR spectroscopy suggest that the dose required to cleave the molecule is no more than 60 mJ/cm$^2$. This high sensitivity to UV light makes conventional pattering using wafer-scale projection lithography feasible. Furthermore, the use of a photosensitive monolayer eliminates the need for complex patterned oxide surfaces, reducing the cost and complexity of device fabrication.

**ACKNOWLEDGMENTS**

The authors thank Dr. Dario Goldfarb for access to the photopatterning equipment. Thanks go to Q. Xu for the access to KPFM.